\documentclass[%
 reprint,
superscriptaddress,
 amsmath,amssymb,aps,
 prl,
]{revtex4-2}

\usepackage{graphicx}
\usepackage{dcolumn}
\usepackage{bm}
\usepackage{xcolor}
\usepackage{amsmath,amssymb}
\usepackage{float}
\usepackage{mathtools}
\usepackage{soul}
\usepackage[T1]{fontenc}
\usepackage{xcolor}
\newcommand\shorthandon{\catcode`@=\active }
\newcommand\shorthandoff{\catcode`@=12 }

\shorthandon
\def@#1@ {\textcolor{red}{#1} }%
\shorthandoff

\begin{document}
\shorthandon

\title{Optical vortex classification via machine learning}

\author{Tobias Schneider}
\affiliation{Department of Physics and Center for Optoelectronics and Photonics Paderborn (CeOPP), Paderborn University, 33098 Paderborn, Germany}

\author{Boqiang Huang}
\affiliation{Department of Mathematics and Computer Science,
University of Cologne, Weyertal 86-90, 50931 Cologne, Germany}

\author{Stefan Schumacher}
\affiliation{Department of Physics and Center for Optoelectronics and Photonics Paderborn (CeOPP), Paderborn University, 33098 Paderborn, Germany}
\affiliation{Institute for Photonic Quantum Systems (PhoQS), Paderborn University, 33098 Paderborn, Germany}
\affiliation{Wyant College of Optical Sciences, University of Arizona, Tucson, Arizona 85721, USA}

\author{Xuekai Ma}
\affiliation{Department of Physics and Center for Optoelectronics and Photonics Paderborn (CeOPP), Paderborn University, 33098 Paderborn, Germany}

\begin{abstract}
Optical vortices carry quantized phase information (topological charge) and are considered candidates for information processing in all-optical circuits. Accurately identifying the quantized vortex charge in a way that is most efficient is essential for data processing. Here, we demonstrate that using only intensity information machine learning algorithms are able to classify vortices into distinct phase categories using a properly trained model. Preprocessing of the original intensity data leads to an improved prediction accuracy of the trained classifier and a much shorter training time ($\sim$3 orders of magnitude faster). The learning mechanism of the algorithms is revealed by the statistical analyses based on calculation of Cohen's $d$. We also find that the training efficiency is associated with nonlinearity, non-Hermiticity, and the vortex shape. In some cases, the trained classifier performs well across distinct physical models. Our findings will benefit and accelerate vorticity-based binary information processing and can also be extended to other physical systems.
\end{abstract}

\maketitle

\section{Introduction}
Machine learning (ML), as the primary engine of artificial intelligence, has also been widely adopted and advanced across scientific disciplines. Physical systems can be used to perform ML algorithms based on the fundamental laws of nature. For example, electronic and neuromorphic systems can mimic the human brain to process information with extreme energy efficiency~\cite{rajendran2019low,kulik2022roadmap}. Optical and photonic systems can process light waves at near-instantaneous speeds with significantly lower energy consumption~\cite{lin2018all,argyris2018photonic,de2019machine,genty2021machine,salmela2021predicting,fu2023photonic,li2023all,ahmed2025universal}. In particular, nonlinearity enables photonic networks to distinguish complex patterns~\cite{opala2022training,zvyagintseva2022machine,sedov2025polariton,gan2025ultrafast,long2025reservoir}. Conversely, ML acting as a high-speed surrogate for expensive experiments and slow mathematical simulations can significantly accelerate data analysis, such as prediction of protein structure~\cite{senior2020improved} and molecular properties~\cite{wu2018moleculenet}, discovery of new materials~\cite{ramprasad2017machine,wei2019machine}, as well as analysis and recognition of cells~\cite{sommer2013machine} and optical signals~\cite{musumeci2018overview}. Classification is a fundamental ML task, with the objective of categorizing disorganized data into predefined groups to enable pattern recognition and automated decision-making. 

\begin{figure}[b]
\centering
{\includegraphics[width=1\linewidth]{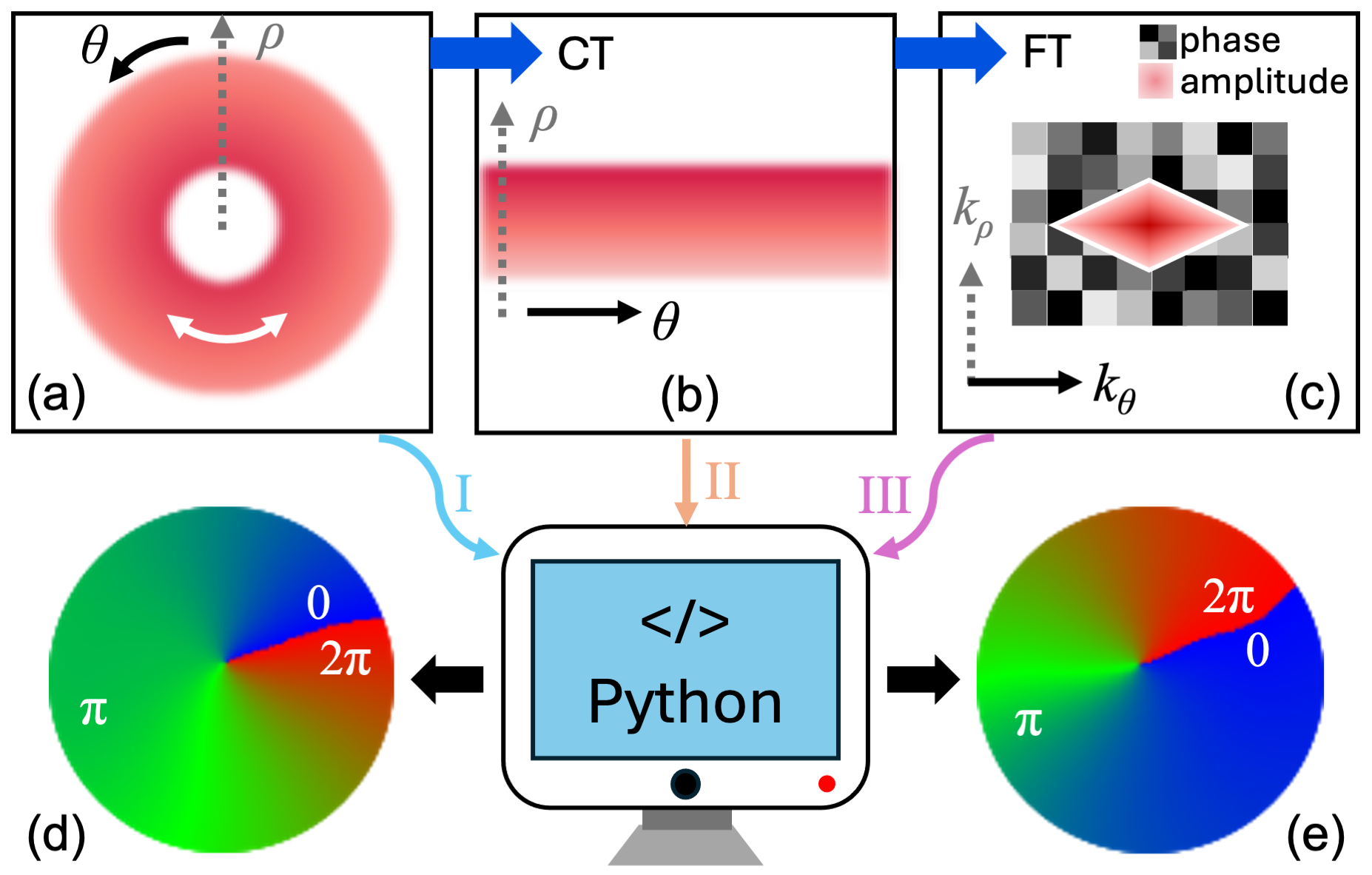}}
\caption{\textbf{The training scheme.} (a) An optical vortex profile in real-space. White arrows indicate possible current-flow directions. (b) Representation of the vortex in (a) after a coordinate transformation (CT) from Cartesian to polar coordinates. (c) $k-$space distribution of the solution in (b) after applying a Fourier transform (FT).Three distinct processes [I: using the original data in (a); II: using the CT data in (b); III: using the FT data in (c)] are performed to train the classifier, respectively, for comparison. The training data are classified into two categories: $m=+1$ (counter-clockwise currents) and $m=-1$ (clockwise currents), corresponding to the phases in (d) and (e), respectively. The trained models are used to classify optical vortices using only intensity information into these two phase categories, thereby identifying the corresponding vorticities.}
\label{fig:1}
\end{figure}

\begin{figure*}[t]
\centering
{\includegraphics[width=1\linewidth]{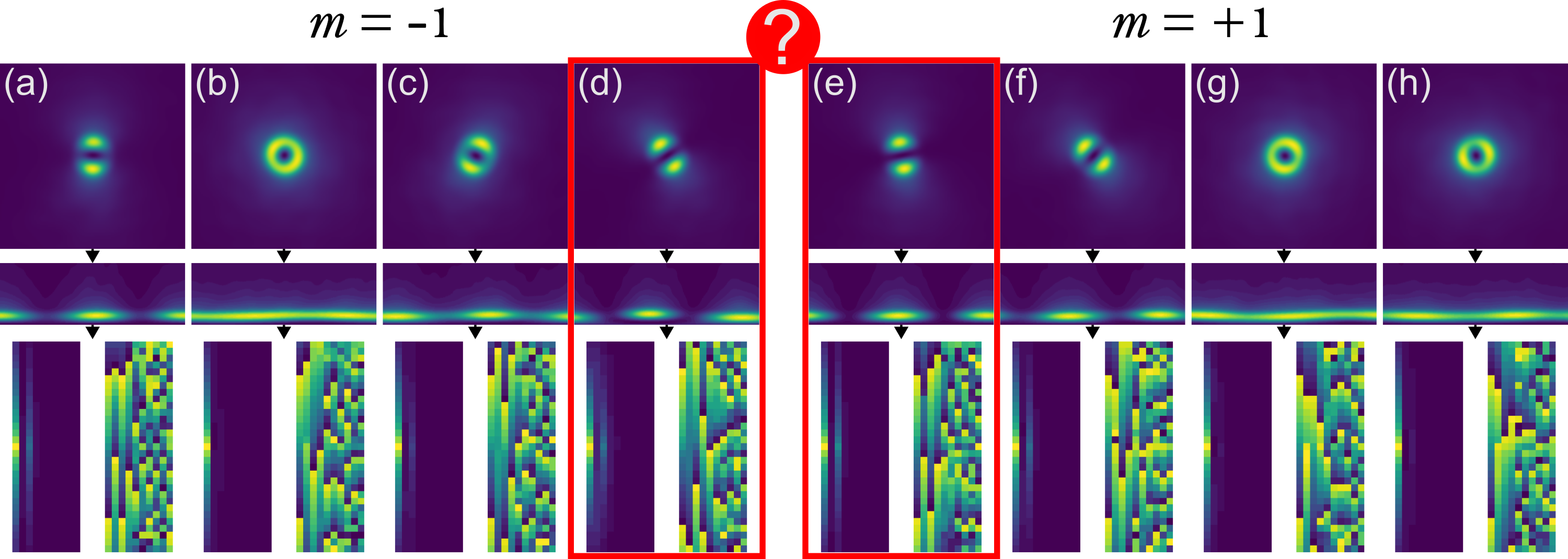}}
\caption{\textbf{Data processing.} Upper: Examples of vortex profiles with (a-d) $m=-1$ and (e-h) $m=+1$. Middle: Corresponding preprocessed CT results, cf. Fig.~\ref{fig:1}(b). Lower: Further preprocessed FT results (Left: amplitude; Right: phase), cf. Fig.~\ref{fig:1}(c). The question mark indicates two profiles that are misclassified by the trained model, while the others can be correctly classified.}
\label{fig:2}
\end{figure*}

A quantum vortex describes the circulation of particles around a singularity, characterized by a quantized phase winding and an associated topological charge (vorticity), which can be utilised for data storage and information processing. An optical vortex can be generated by directly modulating the phase of an optical beam or using metasurfaces~\cite{shen2019optical}. Systems with strong light-matter interaction provide an efficient platform for manipulating orbital angular momentum (OAM) of light, and thus light vorticity~\cite{ma2026vortices}. To determine the topological charge of an optical vortex in experiments, an interferogram with a reference beam is typically measured because the phase information is not directly accessible. In this scenario, a vortex can be identified through a fork shape in the interference pattern, and the fork’s direction indicates the sign of the topological charge, representing clockwise or counterclockwise phase winding. Another efficient method is optical OAM sorting using coordinate transformation~\cite{berkhout2010efficient,ma2020realization}. These two methods, however, normally require additional measurements, complicating vortex readout especially during vortex manipulation~\cite{ma2020realization,PhysRevLett.131.136901,zhai2026observation}.

In this work, we demonstrate the classification of vortex charges using ML algorithms based on  their density information only. Our specific implementation is based on exciton polaritons in planar semiconductor microcavities,  which are hybrid light and matter quasiparticles formed via strong coupling of excitons and photons. We find that for the classifier training, the optimal algorithm is adaptive boosting (AdaBoost) combined with decision trees, which yields a prediction accuracy greater than 95\%. This accuracy can be further increased to more than 99\% by data preprocessing using the Fourier transform. The reason for choosing the exciton polariton system is that it possesses prominent nonlinearity, due to polariton interaction, and a non-equilibrium nature, ascribed to the finite polariton lifetime. In different materials and structures, the polariton interaction and lifetime can be designed on-demand in a very broad range, such that the polariton system can be tuned ranging from the linear to the nonlinear regime and from nearly conservative to highly dissipative. We find that enhanced nonlinearity boosts the prediction accuracy of the trained model by sharpening the statistical differences between the two OAM categories. Conversely, significant vortex ring deformation caused by strong disorder impedes training performance, which scales with the vortex size, as larger vortices encounter more disorder. Interestingly, some of the trained classifiers can predict vortices across different physical models, suggesting the possibility of a foundational model for optical vortex classification and recognition in a broad range of physical systems through further optimization of the training procedure. 

\section{Model Training} 

Figure~\ref{fig:1} illustrates three distinct training processes (Process-I, Process-II, and Process-III) compared in this work. In Process-I, the classifier is trained using only the original intensity/density data of the vortex [Fig.~\ref{fig:1}(a)]. Process-II trains the classifier on data converted from Cartesian to polar coordinates, i.e., a preprocessing procedure applying a coordinate transformation (CT) as shown in Fig.~\ref{fig:1}(a,b). Process-III further processes the CT-transformed data using a Fourier transform (FT), see Fig.~\ref{fig:1}(b,c). In Process-III, additional steps can be performed to further refine the data, such as by removing redundant areas, selecting a smaller window only containing data with relevant densities (higher than $0.1 \%$ of the peak density), or reducing the resolution to 32 points. Once trained, these models predict the topological charges by classifying the input density profiles into the two OAM categories as shown in Fig.~\ref{fig:1}(d,e).

The vortex profiles for training and prediction are numerically generated using the Gross-Pitaevskii model (see Methods), integrated with our numerical solver PHOENIX~\cite{wingenbach2025phoenix}. We generated 2000 samples (1000 each for $m=\pm1$), where random disorder ensures that each profile is unique. Each sample is a steady state in time evolution under the nonresonant excitation of a ring-shaped pump~\cite{ma2020realization} from initial noise. From their density profiles (see Fig.~\ref{fig:2} and more profiles in the SM), one can see that the raw density profiles show no discernible differences that would allow discrimination between their topological charges. While the phase profiles of the FT data exhibit clear variations, they remain visually unclassifiable.

\section{Algorithm Optimization} 
\begin{figure}[t]
\centering
{\includegraphics[width=1\linewidth]{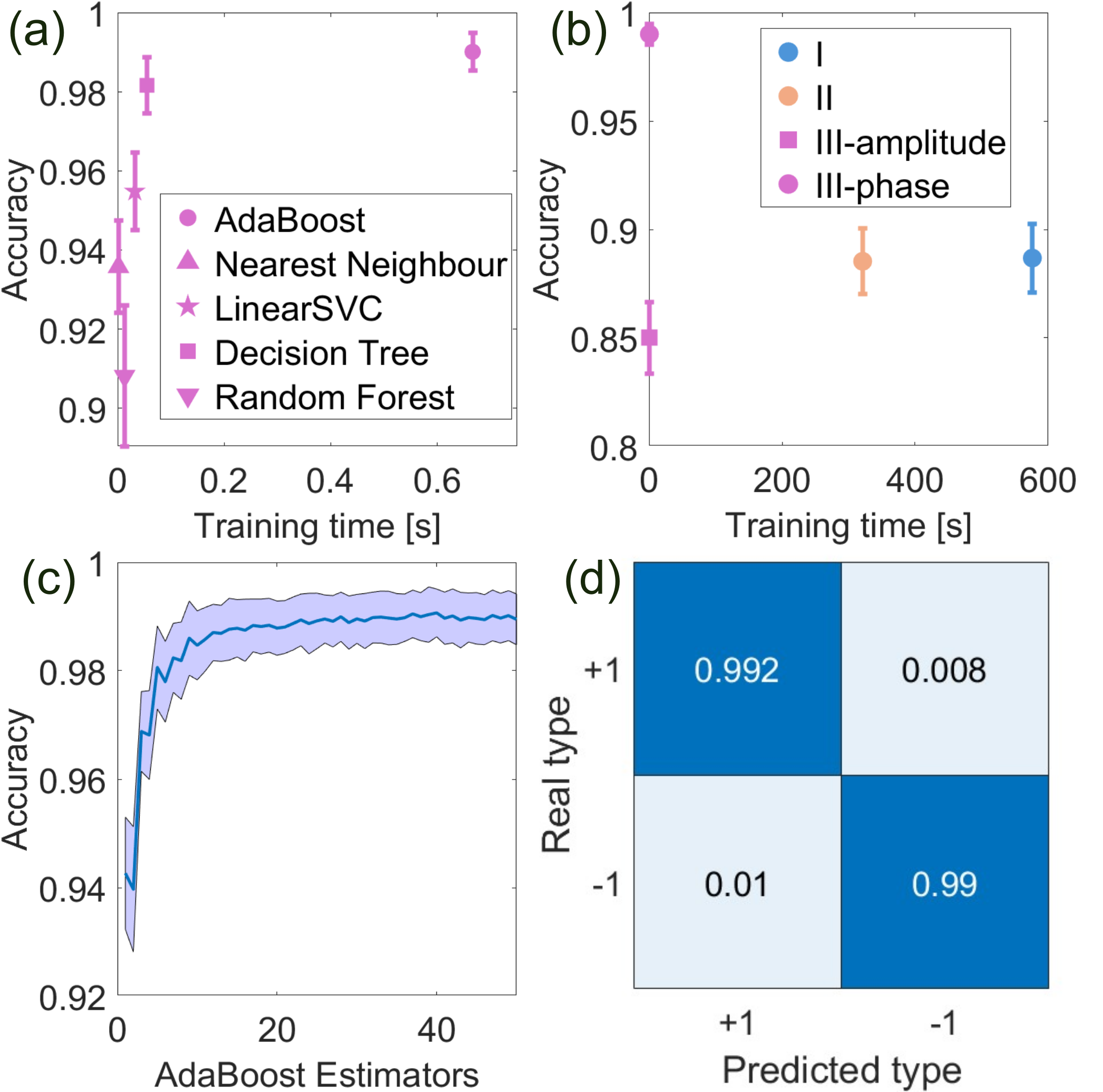}}
\caption{\textbf{Algorithm comparison.} (a) Training time and prediction accuracy of classifiers based on different ML algorithms. (b) Training time and prediction accuracy of AdaBoost with different preprocess steps introduced in Fig.~\ref{fig:1}. Error bars indicate the standard deviation. (c) Dependence of the prediction accuracy on the AdaBoost estimators of process-III-phase. (d) Confusion matrix of the AdaBoost algorithm with process-III-phase.}
\label{fig:3}
\end{figure}

Various ML algorithms can be employed for vortex classification. In this work, we compare several standard algorithms suitable for vortex classification (see Methods for details and codes in GitHub). We evaluate the performance of each algorithm by randomly splitting the 2000 samples into training ($80 \%$) and test ($20 \%$) sets. We repeat this process 200 times to estimate the average accuracy and standard deviation for each model. Figure~\ref{fig:3}(a) summarizes the accuracy and training time for these algorithms. Notably, AdaBoost combined with decision trees achieves the highest accuracy exceeding 99\%, while exhibiting a relatively long training time. These results are based on the phase data (right panels of the lower row in Fig.~\ref{fig:2}) from Process-III (i.e., III-phase). A performance comparison of the three training processes using AdaBoost is shown in Fig.~\ref{fig:3}(b). Although Process-I and Process-II yield accuracies slightly below 90\%, they require significantly more training time ($\sim$ 1000 times longer) than Process-III. For Process-III, both the amplitude (III-amplitude) and phase (III-phase) data allow for rapid training. However, the III-amplitude dataset results in a lower accuracy of roughly 85\%.

Figure~\ref{fig:3}(c) shows the dependence of accuracy and the number of AdaBoost estimators. Accuracy increases rapidly with the estimator count before plateauing at approximately 50. Therefore, we fix the number of AdaBoost estimators at 50 for subsequent models. The corresponding confusion matrix is presented in Fig.~\ref{fig:3}(d). Two misclassified examples, marked by the red frames in Fig.~\ref{fig:2}(d,e), belong to the $m=-1$ and $m=+1$ classes, respectively, but were assigned to the opposite categories. It is worth noting that both samples exhibit severe deformation from the standard ring shape, resembling dipole modes. While all misclassified states share this dipole-like profile, many dipole-like profiles are still classified correctly (see SM).

\begin{figure*}[t]
\centering
{\includegraphics[width=1\linewidth]{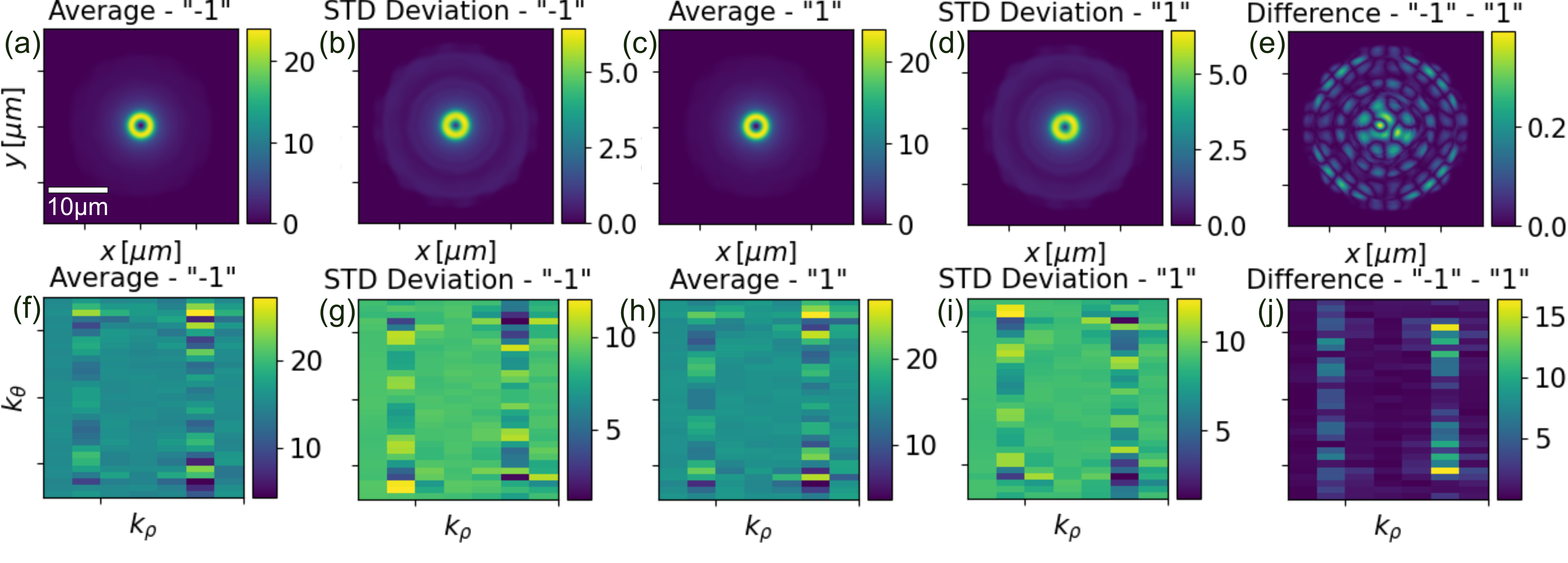}}
\caption{\textbf{Statistical analysis of training data.} Average of the original 1000 vortex profiles of (a) $m=-1$ and (c) $m=+1$, and (b,d) the corresponding standard deviations. Average of the FT phase profiles after preprocessing (i.e., process-III) of (f) $m=-1$ and (h) $m=+1$, and (g,i) the corresponding standard deviations. (e) Absolute difference of the averages in (a) and (c). (j) Absolute difference of the averages in (f) and (h).}
\label{fig:5}
\end{figure*}

To quantify the separability between two vortex categories and investigate the learning mechanism, we calculate Cohen’s $d$, defined as $d_{A,B} = \frac{\overline{S_A} - \overline{S_B}}{\sqrt{\frac{\sigma_A^2 + \sigma_B^2}{2}}}$, where $\overline{S_A}$ and $\overline{S_B}$ denote the mean states, and $\sigma_A$ and $\sigma_B$ represent their respective standard  deviations (STD) for sets A and B. Examples of the corresponding mean profiles and STD deviations for Process-I and Process-III are shown in Fig.~\ref{fig:5}. Comparing the differences between the mean profiles before preprocessing [Fig.~\ref{fig:5}(e)] and after preprocessing [Fig.~\ref{fig:5}(j)] reveals that preprocessing concentrates relevant features into a smaller spatial region while increasing the overall contrast between the states, thereby improving class separability. Since this calculation yields a matrix-valued Cohen's $d$, we define a scalar metric $D$ as the sum of its elements. The resulting $D$ obtained from comparing the $m=+1$ and $m=-1$ classes are shown in Fig. \ref{fig:4} (b), demonstrating a clear positive correlation with classification accuracy.

\section{Role of Physical Properties}
We further investigate the influence of key physical parameters on classification accuracy by varying the nonlinearity, loss rate, ring-pump radius (vortex size), and disorder strength. For each parameter value, a new dataset comprising 2000 samples (1000 per OAM class) is generated, and the model is retrained. The resulting performance is then compared with that obtained using the default dataset.

\begin{figure}[t]
\centering
{\includegraphics[width=1\linewidth]{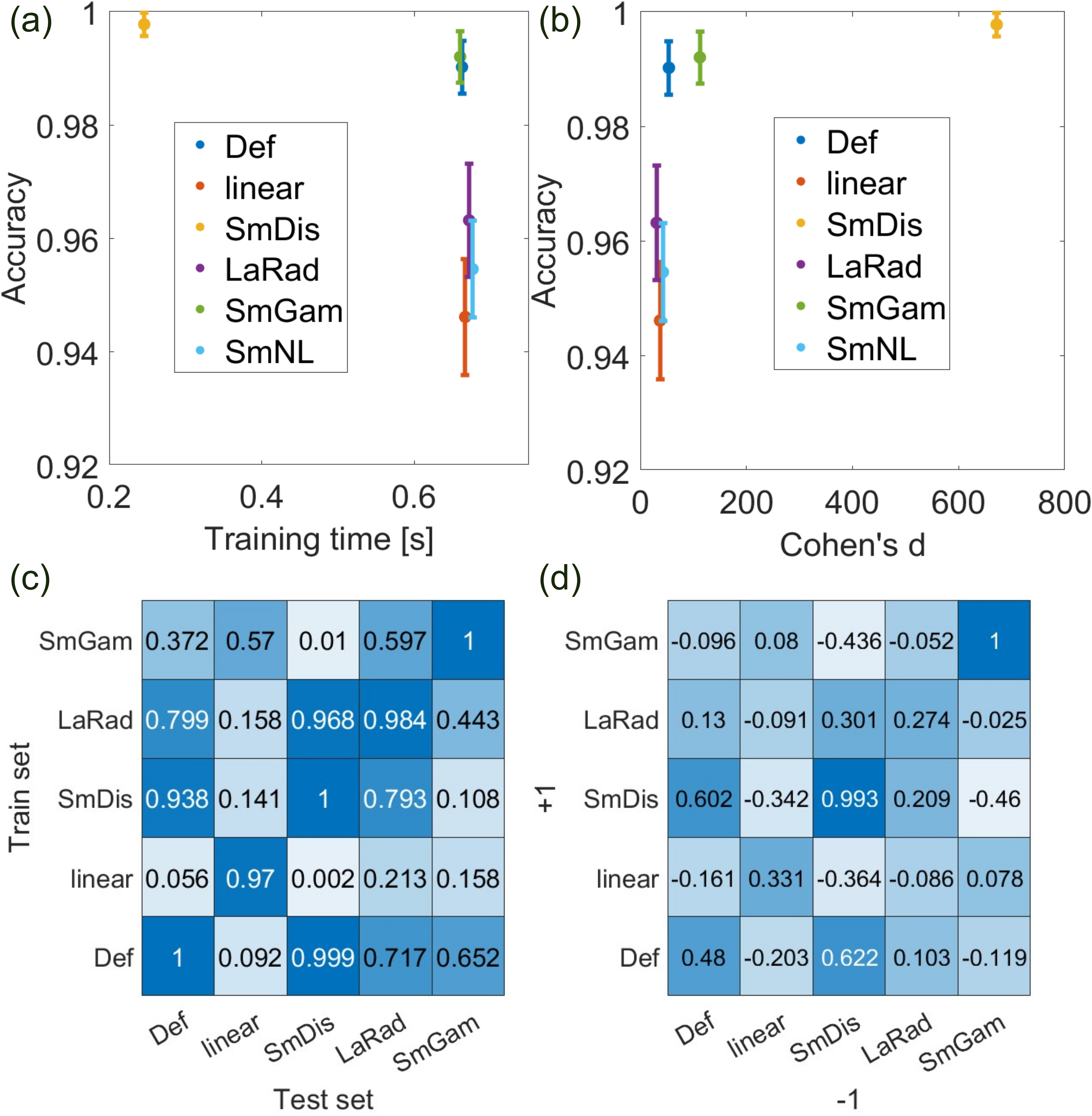}}
\caption{\textbf{Classification performance of distinct physical models.} (a) Training time and prediction accuracy of classifiers trained using different physical models. Def: default case. linear: linear model. SmDis: smaller sample disorder. LaRad: larger vortex radius. SmGam: reduced loss rate $\gamma_c$. SmNL: reduced nonlinearity $g_c$. (b) Cohen's $d$ as a measure of separability between the $m=+1$ and the $m=-1$ sample subsets, shown alongside the corresponding prediction accuracy for each dataset. Error bars indicate the standard deviation. (c) Prediction accuracy of models evaluated on datasets with different physical parameters than those used for training. (d) Difference in Cohen’s $D$ between inter-subset ($m=+1$ vs.\ $m=-1$) and intra-subset (within the same OAM class) comparisons for the different sample sets, normalized to the maximum value.}
\label{fig:4}
\end{figure}

Reducing the nonlinearity strength $g_c$ from the default $10~\mu$eV to half its value ($5~\mu$eV, SmNL) reduces the prediction accuracy from the default $99 \%$ to approximately $96 \%$. In the linear regime with $g_c=0$ (linear, here we focus on only the Kerr nonlinearity), the accuracy decreases to approximately $95 \%$, as shown in Fig.~\ref{fig:4} (a). This drop indicates that intrinsic nonlinearity plays a crucial role in maximizing classification performance. However, the $95 \%$ accuracy remains notably high, demonstrating the model's robustness across both linear and nonlinear regimes.

A similar trend is observed when the radius of the ring pump is increased from $3~\mu$m to $3.5~\mu$m (LaRad), which reduces the accuracy to approximately $96 \%$, see LaRad in Fig.~\ref{fig:4}(a). Notably, the accuracy can be restored to about $98 \%$ by widening the FFT phase selection window in the $k_{\theta}$ direction. This adjustment accounts for higher angular-frequency modes associated with the larger condensate size. However, modifying the preprocessing parameters in this way changes the feature space, thereby limiting direct comparability with the default configuration.

Interestingly, decreasing the disorder scale (SmDis) from $1~\textup{meV}$ to $0.1~\textup{meV}$ improves accuracy to approximately $99.7 \%$ while simultaneously reducing the time required for convergence by approximately $60 \%$, see SmDis in Fig.~\ref{fig:4}(a). Lower disorder reduces intra-class variability, preserving the standard ring-shaped profiles and simplifying the classification task. This increased uniformity is consistent with both the enhanced model performance and significant gains in computational efficiency.

When tuning the system towards a more conservative regime with a smaller condensate loss rate ($\gamma_c$), the accuracy is slightly improved. In general, a smaller loss rate leads to a larger vortex size that should experience a significant background disorder and consequently result in a lower accuracy, see Fig.~\ref{fig:4}(a). However, from the Cohen's $d$ values in Fig.~\ref{fig:4}(b), we find that in a more conservative case, the separability of these two vortex groups becomes higher. 

A classifier trained on a specific parameter set can also be generalized to other parameter sets, as shown in Fig. \ref{fig:4} (c). For example, a model trained on the default dataset maintains high accuracy ($\sim 99.9 \%$) when predicting low-disorder samples, while the reverse yields $\sim 93.8 \%$. Samples generated with a larger ring pump result in reduced cross-set performance, though accuracy remains between $70 \%-80 \%$. The notable exceptions are the linear case with $g_c = 0$ and the more conservative case with $\gamma_c=0.5 ps^{-1}$ , where cross-set accuracy collapses. These models fail to recognize samples produced with non-zero nonlinearity or higher loss rates, with accuracy dropping as low as $\sim 0.2\%$, significantly below the $50 \%$ random baseline. We observe that both the removal of nonlinearity and the reduction of the loss rate drive the system into qualitatively distinct dynamical regimes, thereby complicating direct comparisons and classification compatibility between systems with substantially different values for these parameters.

Another correlation is observed in Fig. \ref{fig:4} (d), where we evaluate a metric comparing the intra-class consistency against the inter-class variance across different parameter sets. This metric, based on the previously defined Cohen’s D, quantifies how well $m=+1$ and $m=-1$ subsets can be distinguished from one another while ensuring samples within the same OAM class remain similar across different physical models. Consequently, larger values indicate stronger cross-set compatibility. As shown in Fig.~\ref{fig:4} (d), this metric correlates strongly with the cross-set accuracies presented in Fig. \ref{fig:4} (c).

\section{Conclusion}
To conclude, we demonstrate that an AdaBoost classifier utilizing decision trees enables highly accurate classification of optical vortex states based on their density profiles (no phase information is needed). Our preprocessing workflow further improves classification performance while substantially reducing training time. We also observe robust cross-compatibility between datasets derived from different physical parameters, suggesting that these models can be generalized well across different systems, with both nonlinearity and loss rate emerging as key factors in model performance. In addition, we show that Cohen's $d$ provides a reliable metric for estimating class separability. Ultimately, a classifier trained on simulated data offers a promising framework for the efficient prediction of vortex states also in experimental settings.

\section*{Acknowledgements}
This work was supported by the Deutsche Forschungsgemeinschaft (DFG) (No. 519608013 and No. 467358803) and the Paderborn Center for Parallel Computing, PC$^2$. 

\bibliography{sample}

\newpage
\section*{Methods}
\subsection{Theoretical Model for Vortex Generation}
The vortices that we study in this work are in exciton-polariton condensates, whose dynamics can be governed by an extended Gross-Pitaevskii (GP) equation coupled to the evolution equation of an exciton reservoir~\cite{PhysRevLett.99.140402}, i.e.,
\begin{equation}
\label{GP_Psi}
\begin{split}
i\hbar \frac{\partial \Psi(\mathbf{r},t)}{\partial t} &= 
\Big[-\frac{\hbar^2}{2m_{\textup{eff}}}\nabla^2 - i\hbar \frac{\gamma_c}{2} + g_c |\Psi(\mathbf{r},t)|^2 \\
&\quad + \big(i\hbar \frac{R}{2} + g_r \big) n + V(\mathbf{r}) \Big] \Psi(\mathbf{r},t)
\end{split}
\end{equation}
\begin{equation}
\begin{aligned}
\label{GP_n}
    \frac{\partial n(\textbf{r},t)}{\partial t} = -(\gamma_r + R |\Psi(\textbf{r},t)|^2) n(\textbf{r},t) + P(\textbf{r},t).
\end{aligned}
\end{equation}
Here, $\Psi(\textbf{r},t)$ is the wavefunction of the polariton condensate with an effective mass $m_{\textup{eff}}=10^{-4} m_\textup{e}$ ($m_\textup{e}$ is the free electron mass). $n(\textbf{r},t)$ is the density of the exciton reservoir. The loss rate of the condensate is $\gamma_c=0.15\ \textup{ps}^{-1}$ and the loss rate of the reservoir is $\gamma_\textup{r}=0.15\ \textup{ps}^{-1}$. The nonlinearity strength is described by $g_\textup{c}$ ($g_\textup{c}=10\ \mu\textup{eV}\ \mu\textup{m}^2$ for the default case), and the interaction of the condensate with the reservoir is $g_\textup{r}=4\ \mu\textup{eV}\ \mu\textup{m}^2$. The condensate is supplemented from the reservoir with a condensation rate $R=0.01\ \textup{ps}^{-1}\ \mu\textup{m}^2$, while the density of the reservoir is generated nonresonantly by an incoherent pump $P(\textbf{r},t)$, which has a ring shape for vortex creation. The potential $V(\textbf{r})$ represents the sample disorder to ensure each vortex profile is unique. To this end, we implement random disorder in each calculation.

\subsection{Model Training Program}
We use the programming language Python together with the scikit-learn library, which provides a broad range of machine learning algorithms. We employ the following classification models: KNeighborsClassifier, SVC, RandomForestClassifier, DecisionTreeClassifier, and AdaBoostClassifier. These models represent a subset of available classifiers, selected based on preliminary benchmarking results. 
The AdaBoostClassifier requires a base estimator; in this work, DecisionTreeClassifier is used, as this combination yielded the most promising results. Specifically, shallow decision trees with a maximum depth of 1 are employed, while the number of estimators is varied (see Fig. 3(c)). 
The preprocessed data are provided to the classifiers as one-dimensional feature vectors, and the corresponding labels are encoded using the built-in LabelEncoder from scikit-learn.

\end{document}